\documentclass[letterpaper]{article} 
\usepackage{aaai2026}  
\usepackage{times}  
\usepackage{helvet}  
\usepackage{courier}  
\usepackage[hyphens]{url}  
\usepackage{graphicx} 
\usepackage{natbib}  
\usepackage{caption} 
\usepackage{algorithm}
\usepackage{algorithmic}

\usepackage[T1]{fontenc}
\usepackage{inconsolata}
\usepackage{subfigure}
\usepackage{enumitem}
\usepackage{booktabs}
\usepackage{multirow}
\usepackage[most]{tcolorbox}
\usepackage{tikz}
\usepackage{algorithm}
\usepackage{algorithmic}
\definecolor{customgray}{RGB}{245,245,245}
\definecolor{customblue}{RGB}{218,232,255}
\definecolor{customgreen}{RGB}{213,232,212}
\definecolor{customred}{RGB}{248,206,204}
\definecolor{custompurple}{RGB}{225,213,231}
\definecolor{customyellow}{RGB}{255,242,204}

\def\methodName{\textsc{CyberRAG}}

\usepackage{newfloat}
\usepackage{listings}
\DeclareCaptionStyle{ruled}{labelfont=normalfont,labelsep=colon,strut=off} 
\floatstyle{ruled}
\newfloat{listing}{tb}{lst}{}
\floatname{listing}{Listing}
\title{Ontology-Aware RAG for Improved Question-Answering in  Cybersecurity Education}
\author{
    Chengshuai Zhao\textsuperscript{\rm 1}
    Garima Agrawal\textsuperscript{\rm 1},
    Fan Zhang\textsuperscript{\rm 1},
    Tharindu Kumarage\textsuperscript{\rm 1},
    Zhen Tan\textsuperscript{\rm 1}, \\
    Yuli Deng\textsuperscript{\rm 1},
    Ying-Chih Chen\textsuperscript{\rm 2}, 
    Huan Liu\textsuperscript{\rm 1}}
\affiliations{
    \textsuperscript{\rm 1}School of Computing and Augmented Intelligence, Arizona State University, USA \\
    \textsuperscript{\rm 2}Mary Lou Fulton Teachers College, Arizona State University, USA\\
    {(czhao93; garima.agrawal; fzhan113; kskumara; ztan36; ydeng19; ychen495; huanliu)}@asu.edu}

\usepackage{bibentry}

\begin{document}

\maketitle

\begin{abstract}
Integrating AI into education has the potential to transform the teaching of science and technology courses, particularly in the field of cybersecurity. AI-driven question-answering (QA) systems can actively manage uncertainty in cybersecurity problem-solving, offering interactive, inquiry-based learning experiences. Recently, Large language models (LLMs) have gained prominence in AI-driven QA systems, enabling advanced language understanding and user engagement. However, they face challenges like hallucinations and limited domain-specific knowledge, which reduce their reliability in educational settings. To address these challenges, we propose \methodName{}, an ontology-aware retrieval-augmented generation (RAG) approach for developing a reliable and safe QA system in cybersecurity education. \methodName{} employs a two-step approach: first, it augments the domain-specific knowledge by retrieving validated cybersecurity documents from a knowledge base to enhance the relevance and accuracy of the response. Second, it mitigates hallucinations and misuse by integrating a knowledge graph ontology to validate the final answer. Comprehensive experiments on publicly available datasets reveal that \methodName{} delivers accurate, reliable responses aligned with domain knowledge, demonstrating the potential of AI tools to enhance education.\footnote{Code:~\url{https://github.com/ChengshuaiZhao0/CyberRAG}}
\end{abstract}

\section{Introduction}
The use of AI in education has the potential to transform the teaching of technology courses~\cite{george2023potential,harry2023role}. In scientific learning, students are expected to engage in problem-solving and exploration, yet traditional classroom methods often focus on the passive acquisition of established knowledge~\cite{ellis1991classroom}. This approach limits opportunities for students to experience the process of knowledge creation, leading to lower cognitive engagement~\cite{corno1983role}. Cybersecurity is a problem-based learning domain where students must master complex tools, develop defense techniques, and uncover new threats, which necessitates a re-imagination of traditional education practices~\cite{shivapurkar2020problem}.

Prior research highlights that managing uncertainty is a crucial component of the learning process, as students often struggle with acquiring new skills, applying diverse methodologies, and forming new understandings~\cite{jordan2015variation}. Educators can effectively manage this uncertainty by increasing it through the introduction of authentic, ambiguous challenges to stimulate critical thinking, maintaining it to encourage deeper exploration and problem-solving, and reducing it by identifying optimal solutions to help students integrate new insights with existing knowledge~\cite{chen2019managing}. AI-driven question-answering (QA) systems can help manage this uncertainty in technical problem-solving by supporting self-paced learning, significantly enhancing cognitive engagement~\cite{means2021hypermodernity}.

In recent years, large language models (LLMs) have become central to AI-driven technologies. While LLM-powered QA systems hold great promise for enhancing learning, they also face challenges such as hallucination and limited domain knowledge, which can undermine their effectiveness~\cite{DBLP:journals/tois/HuangYMZFWCPFQL25,zhao2025chain}. In cybersecurity education, where precision is critical, ensuring the accuracy of AI-generated content is essential. For example, in tasks such as identifying vulnerabilities or interpreting security policies, inaccurate AI responses could lead to misinformation, compromising the learning experience and, potentially, real-world security~\cite{kumarage2024survey}. A promising solution to address this challenge is the retrieval-augmented generation (RAG) approach, where the model generates responses by retrieving information from a validated knowledge base, thereby enhancing the accuracy and reliability~\cite{DBLP:conf/nips/LewisPPPKGKLYR020}.

Although the RAG approach helps reduce hallucinations and address domain knowledge issues to some extent, the reliability of LLM-generated answers remains a concern for achieving educational goals. Students may ask questions that fall outside the scope of the augmented cybersecurity knowledge base. In such cases, LLMs rely on their own parametric knowledge to generate responses, which can expose the QA system to risks of misinformation or misuse~\cite{DBLP:journals/corr/abs-2402-14859, DBLP:conf/emnlp/TanZMLWLCL24}. In an educational setting, it is also crucial to prevent students from manipulating the AI system for unintended purposes. There is a strong need to provide a validation system to ensure the accuracy and safety of LLM-generated responses. One potential solution is safety alignment~\cite{DBLP:conf/nips/ChristianoLBMLA17}. However, this method requires verification by cybersecurity experts, making it labor-intensive, costly, and time-consuming. Preferably, an automatic validation approach is needed. Domain-specific knowledge graphs, which structure expert knowledge and capture the interactions between key entities in alignment with domain rules~\cite{abu2021domain}, offer a promising direction. The knowledge graph ontology encodes these rules~\cite{kejriwal2019domain}. By leveraging this ontology, LLM responses can be validated by fact-checking against the predefined rules, ensuring greater accuracy and reliability without the need for constant human oversight.

In this paper, we propose \textbf{\methodName{}}, an ontology-aware RAG approach for developing a reliable QA system in cybersecurity education, comprising two key components. First, we utilize RAG methods to retrieve validated cybersecurity documents from a knowledge base, enhancing both the accuracy and relevance of the answers. Through finely crafted prompts, \methodName{} improves QA performance by leveraging both cybersecurity content and the natural language processing capabilities of LLMs. Additionally, we introduce an ontology-based validation approach that uses a cybersecurity knowledge graph ontology to automatically verify LLM-generated responses, thereby preventing potential risks of misuse and hallucination. We conduct comprehensive experiments on publicly available datasets to demonstrate the effectiveness of \methodName{} in delivering reliable and accurate answers. Our research explores the potential of integrating AI into education, emphasizing its transformative impact on traditional methods. This approach extends beyond cybersecurity and can also be applied to other educational subjects. Our contributions can be summarized as follows:

\begin{itemize}
    \item \textbf{Ontology-aware RAG for cybersecurity QA.} We introduce \methodName{}, a plug-and-play pipeline that couples dense retrieval and LLM generation with an ontology-based validator instantiated from AISecKG, yielding accurate and safe answers for problem-based learning.
    \item \textbf{Fine-grained prompting for accuracy and safety.} We design reusable prompts for answer generation and ontology-aligned validation that operationalize domain rules as checks, automatically filtering hallucinations and rejecting out-of-scope or misuse-seeking queries.
    \item \textbf{Comprehensive evaluation and ablations.} We evaluate \methodName{} on CyberQ across \emph{In-KB}, \emph{Out-of-KB}, and \emph{Zero-shot} settings with multiple LLM backbones and retrievers, using n-gram and RAGAS metrics. Ablations vary KB coverage, prompting strategies, and retriever choice, and we analyze an ontology gate on mixtures with non-cybersecurity QA sets.
    \item \textbf{Actionable guidance for deployment.} From these studies, we provide actionable deployment guidance by clarifying how KB coverage, retrieval strategies, and model/prompting decisions shape quality, safety, and efficiency, leading to a transparent, auditable QA pipeline with applications beyond cybersecurity education
\end{itemize}

\section{Related Work}
\subsection{RAG in Education}
Generative models have the potential to transform traditional education by enabling personalized learning and automating content creation, making education more adaptive and accessible~\cite{george2023potential}. The integration of LLMs in education has further enhanced conversational capabilities, supporting self-paced learning through AI-driven question-answering tools and bots~\cite{moore2023empowering}. However, the use of LLMs also presents challenges, particularly the risks of hallucinations and inaccurate responses, which are critical concerns in educational contexts and must be carefully managed to ensure responsible implementation~\cite{yan2024practical}. To address these issues, retrieval-augmented generation (RAG) methods are used by combining LLMs with real-time and up-to-date knowledge base retrieval to improve response accuracy~\cite{liu2024hita}. However, relying solely on RAG is insufficient to ensure reliability in educational settings.

\subsection{LLM-generated Answer Validation}
However, most LLM approaches are currently limited to handling simple queries, as generating complex answers from extensive knowledge articles and course materials sourced from diverse knowledge bases remains challenging, often leading to issues with the correctness of the answers~\cite{elmessiry2024navigating}. Jeong et al.~\cite{jeong2024adaptive} proposed an adaptive QA approach to handle complex questions, while Li et al.~\cite{li2024grammar} developed an evaluation framework for grounded and modular assessment of RAG responses.  Various domains are using knowledge graphs(KGs) to enhance LLMs' reasoning and retrieval capabilities~\cite{hussien2024rag}. An ontology within a KG, which defines the rules of a specific domain, can also be used to validate LLM responses by checking the correctness of relationships between key entities. This capability is particularly valuable in educational settings, where ensuring the accuracy of LLM-generated content is crucial. While some student questions may extend beyond the scope of the augmented course material or knowledge base, ontology-aware validation offers a promising solution. However, research in this area remains limited. Our ontology-aware methods mark a significant advancement in this field.

\subsection{AI for Cybersecurity Education} 
Learning cybersecurity is crucial for national security, safeguarding critical infrastructure, and ensuring defense~\cite{aldaajeh2022role, newhouse2017national}. However, mastering this complex field requires deep knowledge of concepts, tools, and attack-defense simulations~\cite{cheung2011challenge}. While AI shows promising potential in enhancing cognitive learning, research on its application in cybersecurity education remains limited~\cite{wei2023cybersecurity}. Previous efforts include structured flow graphs from Capture The Flag (CTF) texts for vulnerability analysis~\cite{pal2021constructing} and knowledge graphs for guiding student projects~\cite{deng2019knowledge}. However, a significant gap remains: the lack of a self-paced cybersecurity QA framework to enable interactive learning for students. Our approach leverages an ontology-aware RAG, offering an effective and reliable solution for cybersecurity education.

\begin{figure*}[t]
    \centering
    \includegraphics[width=\linewidth]{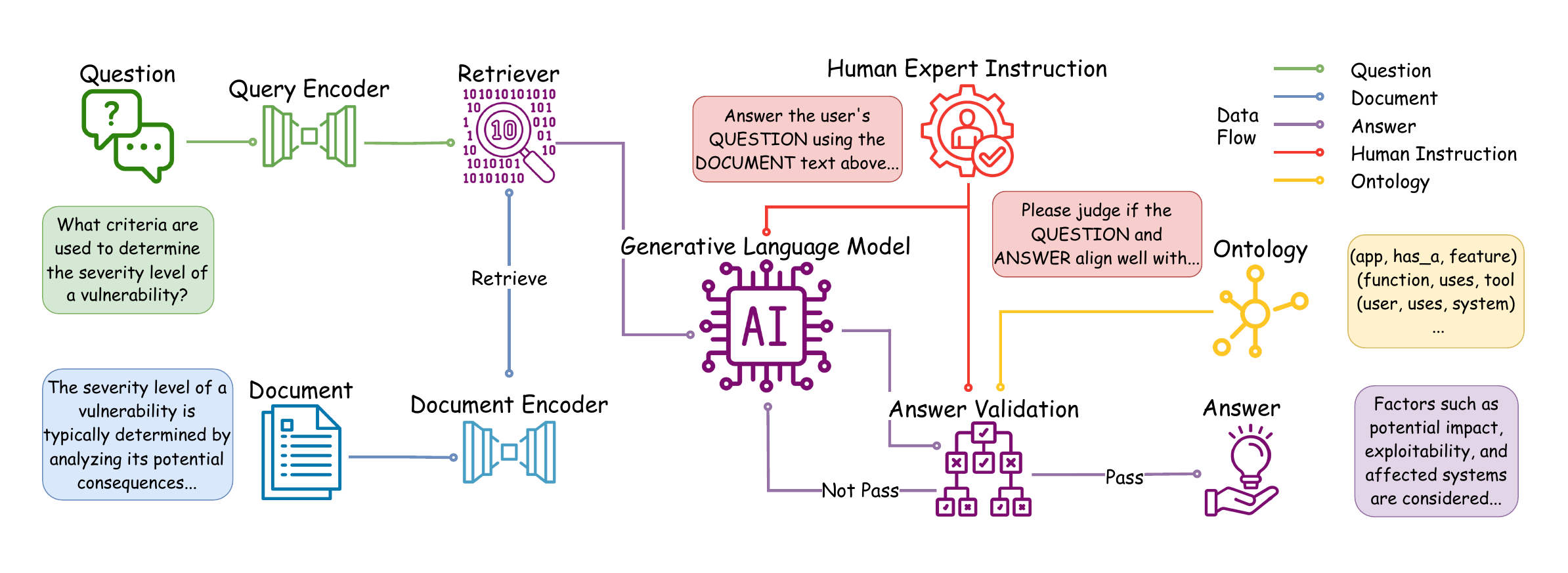}
    \caption{Overview of \methodName{} Framework.}
    \label{fig:overview}
\end{figure*}

\section{Preliminaries}
\subsection{Problem Formulation}
We consider cybersecurity problem-based learning to be a question-answering (QA) problem. Specifically, students raise a series of questions $Q = \{q_1, q_2,..., q_{|Q|}\}$, we expect the model to output corresponding responses as answers $A = \{a_1, a_2,..., a_{|A|}\}$, where $|Q| = |A|$ in our scenario.

\subsection{RAG System}
We employ a retrieval-augmented generation (RAG) system as a base framework to solve our problem because of its impressive capacity for hallucination mitigation. A classic RAG approach entails two parts:
\begin{itemize}
    \item  A retriever $\mathcal{R}$ that can effectively retrieve reference documents $D = \{d_1, d_2,..., d_{|D|}\}$ from the knowledge base $K$ (e.g., knowledge graph, QA database, and text materials) based on query-document relevance.
    \item A generation model (e.g., large language model) $\mathcal{G}$ that provides answers using the documents and human instructions $I = \{i_1, i_2,..., i_{|I|}\}$.
\end{itemize}

\subsection{Knowledge Graph Ontology}
In the context of our \methodName{} framework, we incorporate knowledge graph ontology to validate the LLM-generated responses. A knowledge graph (KG) is a structured representation of knowledge in which entities (a.k.a, nodes) are connected by relationships (a.k.a, edges). Formally, we define a knowledge graph as a tuple $G = (E, R)$, where $E = \{e_1, e_2, ..., e_{|E|}\}$ represents the set of entities, and $R = \{r_1, r_2, ..., r_{|R|}\}$ represents the set of relationships between these entities. Each relationship $r \in R$ can be viewed as a directed edge connecting two entities $e_i$ and $e_j$ in the graph, thereby forming a triplet $(e_i, r, e_j)$.

To further formalize the structure and semantics of the domain knowledge represented in our system, we utilize an ontology $O$. An ontology is a formal specification of a set of concepts within a domain and the relationships between those concepts. Formally, we define an ontology $O$ as a tuple $O = (C, R_C, H_C)$, where $C = \{c_1, c_2, ..., c_{|C|}\}$ is a set of concepts, $R_C = \{r_1^C, r_2^C, ..., r_{|R_C|}^C\}$ is a set of conceptual relationships, and $H_C \subseteq C \times C$ is a hierarchical structure (e.g., a taxonomy) that organizes these concepts.

\section{The proposed CyberRAG}
Our proposed framework, named \textbf{\methodName{}}, comprises two key components: a retrieval-augmented generation (RAG) system and an ontology-based answer validation module. By leveraging the RAG system with carefully crafted prompts, \methodName{} effectively addresses students' questions by utilizing cybersecurity knowledge materials and the natural language understanding capabilities of LLMs. The ontology-based validation approach ensures that the responses provided to students are both reliable and secure. An overview of \methodName{} is presented in Fig.~\ref{fig:overview}.

\subsection{Cybersecurity Knowledge Retrieval}
\label{sec:retriever}
An effective retriever is essential for aligning student queries with relevant cybersecurity materials. We adopt a dual-encoder dense retrieval architecture consisting of a \textit{question encoder} $\mathcal{E}_Q$ and a \textit{document encoder} $\mathcal{E}_D$. The question encoder projects a student query $q_i \in Q$ into a semantic vector $h_{q_i} = \mathcal{E}_Q(q_i)$, while the document encoder maps a knowledge base document $d_j \in D$ into $h_{d_j} = \mathcal{E}_D(d_j)$. These encoders are trained to preserve semantic proximity, ensuring that queries and relevant documents are embedded close to each other in the latent space. In practice, the same encoder can be shared between questions and documents, depending on the setting.

The semantic similarity between query and document representations is measured using cosine similarity: $sim(h_{q_i}, h_{d_j}) = \frac{h_{q_i} \cdot h_{d_j}}{\|h_{q_i}\|\|h_{d_j}\|}$. Based on this score, the retriever selects the top-$k$ documents $\mathcal{R}(Q, k)$ most relevant to the query. These retrieved materials form the knowledge context for subsequent answer generation.

\subsection{Cybersecurity Answer Generation}
After retrieving relevant cybersecurity documents, the next step is to prompt the generative model (i.e., LLMs) to generate answers. The most challenging aspect is designing a prompt that maximizes the effectiveness of both the retrieved information and the generative model. An ideal prompt should guide the model to summarize the information when the retrieved documents are highly relevant and encourage the model to generate new responses by leveraging its own knowledge when the retrieved information is insufficient or incomplete.

\tcbset{
    promptstyle/.style={
        colback=gray!5,
        colframe=black,
        fonttitle=\bfseries,
        boxrule=0.5mm,
        sharp corners,
        enhanced,
        breakable,
        width=1\linewidth
    }
}

\begin{tcolorbox}[promptstyle, title=Answer Generation Prompt]
\small
\ttfamily
DOCUMENT:\\
\{document\}\\

QUESTION:\\
\{question\}\\

INSTRUCTIONS:\\
Answer the user's QUESTION using the DOCUMENT text above. Keep your answer grounded in the facts of the DOCUMENT. If the DOCUMENT doesn’t contain the facts to answer the QUESTION, give a response based on your knowledge.
\end{tcolorbox}

We design the prompt as illustrated in the \textbf{Answer Generation Prompt}. The proposed prompt consists of three key elements: (i)~Documents: the documents relevant to the student's query are retrieved by the retriever from the knowledge base, using the method outlined in the previous section. (ii)~Questions: these are queries raised by students, which the generative model is tasked with answering. (iii)~Instructions: These are rules and prompts designed by domain experts. The first two instructions guide the generative model to answer students' questions by referencing the retrieved documents. The third instruction encourages the model to generate responses based on its own knowledge when the documents are insufficient. Thus, the proposed framework can address students' queries, whether or not they are covered by the knowledge base.
The final prompt is given as input to the generative model to generate the answer to the student's query: $A = \mathcal{G}(D, Q, I)$.

\subsection{Ontology-based Answer Validation}
For ontology validation, we first extract and distill ontology from AISecKG~\cite{agrawal2023aiseckg}, a cybersecurity knowledge graph that defines relationships between concepts, applications, and roles within the cybersecurity domain. And the ontology is further validated and evaluated by three domain experts. Finally, we acquire ontology with three broad categories and 12 entity types. Concepts include features, functions, data, attacks, vulnerabilities, and techniques, while applications cover tools, systems, and apps. Roles consist of users, attackers, and security teams. The ontology defines nine core relationships between these entities, represented by 68 unique edges. For example, tuples such as (`attacker', `can\_exploit', `feature') and (`security team', `can\_analyze', `feature') illustrate entities and these relationships. These triples represent the fundamental domain rules that govern cybersecurity information at a schema level. By leveraging these ontology-based triples along with the natural language understanding capabilities of LLMs, an automatic answer validation system can be developed. 

Specifically, given the question $Q$ and the corresponding response $A$ provided by the generative language model, the validation model $\mathcal{V}$ takes the QA contexts, ontology rules denoted by $O = \{o_1, o_2,..., o_{|O|}\}$, and validation human instruct $I'$ as the input, then produces the validation result $R$: $R = \mathcal{V}(Q, A, O, I')$. Note that $R \in [0, 1]$, and a higher score indicates the response aligns with the ontology well. By setting an appropriate threshold~$\sigma$, the pipeline can effectively filter out incorrect answers and identify potential misuse behavior. The prompt is shown in the \textbf{Ontology Validation Prompt}.

\begin{tcolorbox}[promptstyle, title=Ontology Validation Prompt]
\small
\ttfamily
QUESTION:\\
\{question\}\\

ANSWER:\\
\{answer\}\\

ONTOLOGY:\\
\{ontology\}\\

INSTRUCTIONS:\\
Please judge if the QUESTION and ANSWER align well with the ONTOLOGY. The QUESTION and ANSWER align well with the ONTOLOGY if they are in the same knowledge domain as the ONTOLOGY, and the ANSWER follows the relationships defined in the ONTOLOGY.
\end{tcolorbox}



\section{Experiment and Discussion}
\subsection{Dataset}
In this work, we use CyberQ~\cite{agrawal2024cyberq}, an open-source cybersecurity dataset containing around 4,000 open-ended questions and answers on topics such as cybersecurity concepts, tool usage, setup instructions, attack analysis, and defense strategy. The CyberQ dataset includes three sub-datasets: (i) \textbf{CyberQ-ZS} contains 1,027 QA pairs for simple WH-questions on cybersecurity entities. (ii) \textbf{CyberQ-FS} consists of 332 medium-complexity QA pairs related to setup and tools. (iii) \textbf{CyberQ-OD} includes 2,171 high-complexity QA pairs covering attack and defense scenarios. Collectively, these datasets provide a comprehensive and challenging overview of cybersecurity knowledge, making them ideal for our goal of developing an interactive QA system to teach cybersecurity to students.

\begin{table*}[!th]
\centering
\caption{Evaluation of \methodName{} with various large language models backbones across scenarios for CyberQ}
\label{tab:main_results}
\setlength{\tabcolsep}{3pt}%
\renewcommand{\arraystretch}{1.3}%
\resizebox{\linewidth}{!}{
\begin{tabular}{l*{12}{c}}
\toprule
\textbf{Model~$\downarrow$} & \multicolumn{4}{c}{\textbf{CyberQ-ZS}} & \multicolumn{4}{c}{\textbf{CyberQ-FS}} & \multicolumn{4}{c}{\textbf{CyberQ-OD}} \\
\cmidrule(lr){2-5}\cmidrule(lr){6-9}\cmidrule(lr){10-13}
\textbf{Metric~$\to$} & \textbf{BERTScore~$\uparrow$} & \textbf{METEOR~$\uparrow$} & \textbf{ROUGE-1~$\uparrow$} & \textbf{ROUGE-2~$\uparrow$} 
& \textbf{BERTScore~$\uparrow$} & \textbf{METEOR~$\uparrow$} & \textbf{ROUGE-1~$\uparrow$} & \textbf{ROUGE-2~$\uparrow$} 
& \textbf{BERTScore~$\uparrow$} & \textbf{METEOR~$\uparrow$} & \textbf{ROUGE-1~$\uparrow$} & \textbf{ROUGE-2~$\uparrow$} \\
\midrule
\multicolumn{13}{c}{\textit{In KB}} \\
\midrule
Llama-2-7b   & 0.920 & 0.670 & 0.518 & 0.486 & 0.942 & 0.783 & 0.682 & 0.639 & 0.920 & 0.659 & 0.488 & 0.454 \\
Llama-2-13b  & 0.899 & 0.587 & 0.406 & 0.373 & 0.928 & 0.741 & 0.642 & 0.604 & 0.908 & 0.615 & 0.440 & 0.406 \\
Llama-3-8B   & \underline{0.925} & \underline{0.773} & \underline{0.630} & \underline{0.584} 
             & \underline{0.946} & \underline{0.856} & \underline{0.785} & \underline{0.717} 
             & \underline{0.929} & \underline{0.763} & \underline{0.614} & \underline{0.564} \\
Mistral-7B   & \textbf{0.963} & \textbf{0.916} & \textbf{0.879} & \textbf{0.828} 
             & \textbf{0.982} & \textbf{0.950} & \textbf{0.940} & \textbf{0.909} 
             & \textbf{0.968} & \textbf{0.918} & \textbf{0.880} & \textbf{0.827} \\
Qwen2.5-7B   & 0.885 & 0.584 & 0.363 & 0.310 & 0.914 & 0.722 & 0.546 & 0.461 & 0.896 & 0.617 & 0.409 & 0.336 \\
Qwen3-8B     & 0.864 & 0.470 & 0.193 & 0.181 & 0.880 & 0.554 & 0.304 & 0.295 & 0.868 & 0.494 & 0.216 & 0.205 \\
\midrule
\multicolumn{13}{c}{\textit{Out of KB}} \\
\midrule
Llama-2-7b   & 0.867 & 0.324 & 0.190 & 0.091 & 0.871 & 0.335 & 0.274 & 0.127 & 0.873 & 0.360 & 0.220 & 0.105 \\
Llama-2-13b  & 0.866 & 0.331 & 0.203 & 0.094 & 0.872 & 0.361 & 0.318 & 0.146 & 0.874 & 0.371 & 0.241 & 0.114 \\
Llama-3-8B   & 0.865 & 0.337 & 0.204 & 0.095 & 0.872 & 0.359 & 0.301 & 0.144 & 0.874 & 0.380 & 0.248 & 0.117 \\
Mistral-7B   & 0.883 & 0.365 & 0.295 & 0.134 & 0.884 & 0.334 & 0.363 & 0.172 & 0.893 & 0.411 & 0.344 & 0.163 \\
Qwen2.5-7B   & 0.854 & 0.301 & 0.177 & 0.078 & 0.861 & 0.339 & 0.278 & 0.121 & 0.857 & 0.337 & 0.203 & 0.088 \\
Qwen3-8B     & 0.840 & 0.254 & 0.117 & 0.049 & 0.844 & 0.291 & 0.192 & 0.079 & 0.846 & 0.291 & 0.146 & 0.062 \\
\midrule
\multicolumn{13}{c}{\textit{Zero Shot}} \\
\midrule
Llama-2-7b   & 0.867 & 0.313 & 0.180 & 0.087 & 0.873 & 0.348 & 0.245 & 0.109 & 0.869 & 0.323 & 0.188 & 0.088 \\
Llama-2-13b  & 0.869 & 0.315 & 0.177 & 0.086 & 0.875 & 0.355 & 0.245 & 0.111 & 0.871 & 0.326 & 0.188 & 0.089 \\
Llama-3-8B   & 0.842 & 0.277 & 0.161 & 0.079 & 0.857 & 0.341 & 0.238 & 0.109 & 0.850 & 0.300 & 0.177 & 0.086 \\
Mistral-7B   & 0.869 & 0.323 & 0.191 & 0.088 & 0.876 & 0.368 & 0.272 & 0.119 & 0.870 & 0.330 & 0.204 & 0.089 \\
Qwen2.5-7B   & 0.862 & 0.293 & 0.155 & 0.076 & 0.869 & 0.355 & 0.227 & 0.106 & 0.864 & 0.311 & 0.169 & 0.081 \\
Qwen3-8B     & 0.859 & 0.280 & 0.153 & 0.073 & 0.873 & 0.361 & 0.253 & 0.118 & 0.865 & 0.313 & 0.181 & 0.084 \\
\bottomrule
\multicolumn{13}{c}{\small Each result is the mean of ten runs, with standard deviations under $10^{-3}$ across all experiments.}
\end{tabular}}
\end{table*}

\subsection{Experiment and Parameter Settings}
\textbf{Knowledge Base.} Our knowledge base comprises course materials and laboratory manuals from graduate-level cybersecurity courses, with AISecKG~\cite{agrawal2023aiseckg} serving as the underlying cybersecurity knowledge graph for ontological validation.

\textbf{Experimental Scenarios.} We evaluate \methodName{} across three distinct scenarios: (1) \textbf{In-KB}: queries answerable using existing knowledge base documents; (2) \textbf{Out-of-KB}: queries requiring information not present in the knowledge base; and (3) \textbf{Zero-shot}: baseline configuration without knowledge base augmentation, relying solely on LLM parametric knowledge.

\textbf{Model Configurations.} For document retrieval, we employ four state-of-the-art dense retrievers: Contriever~\cite{DBLP:journals/tmlr/IzacardCHRBJG22}, Contriever-MS~\cite{DBLP:journals/tmlr/IzacardCHRBJG22}, DPR-NQ~\cite{karpukhin2020dense}, and DPR-MultiQA~\cite{karpukhin2020dense}, utilizing cosine similarity for semantic relevance scoring. Our generative models include Llama-2-7B/13B-Chat~\cite{touvron2023llama}, LLaMA3-8B-Instruct~\cite{DBLP:journals/corr/abs-2407-21783}, Mistral-7B-Instruct~\cite{Jiang2023Mistral7}, Qwen2.5-7B-Instruct~\cite{Yang2024Qwen25TR}, and Qwen3-8B~\cite{yang2025qwen3}. The ontology validation component is implemented using LLaMA3-8B-Instruct~\cite{DBLP:journals/corr/abs-2407-21783}.

\textbf{Hyperparameters.} We set the context window to 1,024 tokens for all language models, employ beam search with 4 beams during generation. Unless otherwise specified, Contriever and LLaMA3-8B-Instruct are used. All unspecified parameters follow default model configurations.

\textbf{Evaluation Metrics.} We assess system performance using four established metrics: BERTScore~\cite{DBLP:conf/iclr/ZhangKWWA20}, METEOR~\cite{DBLP:conf/acl/BanerjeeL05}, ROUGE-1~\cite{lin2004rouge}, and ROUGE-2~\cite{lin2004rouge}. Ground truth annotations are provided by cybersecurity domain experts from the CyberQ dataset. All experiments are conducted over 10 independent runs, reporting mean scores with standard deviations.

\subsection{Superior Performance of Proposed CyberRAG}
To answer the question of how effectively proposed \methodName{} facilitates question-answering for cybersecurity education, we conduct comparative experiments by replacing the backbone LLM in \methodName{} with several alternatives and evaluating across three scenarios summarized in Table~\ref{tab:main_results}.
  
We can observe that \methodName{} consistently delivers competitive results. For example, in the In-KB CyberQ-FS, LLaMA-2-13B achieves a BERTScore of 0.928, METEOR of 0.741, ROUGE-1 of 0.642, and ROUGE-2 of 0.604, demonstrating both semantic alignment and lexical fidelity with expert answers. Even smaller backbones, such as \textit{Qwen2.5-7B}, benefit substantially from retrieval support, achieving stronger In-KB performance with a BERTScore of 0.914.

Across all backbones, performance is higher for In-KB queries than Out-of-KB ones. For instance, LLaMA-2-7B records a BERTScore of 0.920 on In-KB CyberQ-OD queries, but only 0.873 in Out-of-KB FS queries. Similarly, Qwen3-8B improves from just 0.844 in Out-of-KB CyberQ-FS to 0.880 in In-KB settings. This demonstrates that grounding responses in the knowledge base provides reliable contexts that boosts model accuracy across both simple and complex queries.  

Even when relevant documents are absent, \methodName{} produces meaningful answers. For example, in the Out-of-KB CyberQ-ZS subset, LLaMA-3-8B achieves a BERTScore of 0.865 with METEOR 0.337 and ROUGE-1 0.204. Likewise, Qwen2.5-7B produces a BERTScore of 0.857 on the OD subset, showing that although wording may diverge, the responses still capture the essence of ground-truth expert annotations.  

Different backbones show distinct behaviors. Mistral-7B dominates overall with near-human In-KB performance (e.g., ROUGE-1 above 0.94 in FS). LLaMA-3-8B provides a strong balance across In-KB and Out-of-KB tasks. LLaMA-2 models are moderate but stable, while Qwen models underperform in both scenarios, especially Qwen3-8B with an In-KB BERTScore of only 0.868 on CyberQ0-ZS dataset, suggesting weaker adaptability to cybersecurity knowledge. This diversity highlights how \methodName{} is robust enough to improve weaker backbones while maximizing stronger ones.

Together, these findings confirm that CyberRAG significantly improves QA performance by combining retrieved documents and LLM-based generation. While Mistral-7B stands out as the best backbone, the framework ensures that even weaker backbones, such as Qwen2.5-7B and Qwen3-8B, benefit meaningfully. The substantial boost in In-KB scenarios underscores CyberRAG’s potential as an effective and interactive assistant for cybersecurity education.

\subsection{Ablation Study}
We conducted an ablation study to investigate the individual contribution of each component to \methodName{}'s overall performance. We consider two different scenarios: (i) \methodName{} is built solely on a question-answering dataset, such as CyberQ, without utilizing the generative language model. In this case, the QA system is unable to answer questions beyond the scope of the training dataset, which is undesirable as it limits the breadth of cybersecurity education. (ii) \methodName{} without augmenting the knowledge base. In this case, the model essentially reverts to functioning as a generative language model, answering questions in a zero-shot manner. As shown in Fig.~\ref{tab:main_results}, the results under zero-shot settings are considerably inferior. Compared to the In-KB scenarios, the model under the zero-shot setting lacks access to the ground truth information for reference, leading to a significant drop in performance.
When comparing the zero-shot results with Out-of-KB scenarios, the latter demonstrates a better performance. This is because, although Out-of-kb can not be answered directly using documents from the knowledge base, these documents still provide relevant and supplementary information that may help in formulating the answers. In conclusion, both the knowledge base and the generative language model play significant roles in enhancing the performance of a QA system in cybersecurity education.

\subsection{RAG Analysis}
To evaluate the effectiveness of retrieval-augmented generation (RAG) in improving answer quality and mitigating hallucinations, we conduct a comprehensive RAG analysis. Specifically, we examine the documents retrieved from the knowledge base and compare the responses generated by \methodName{} against the ground truth to assess the contribution of the retrieval process. We leverage RAGAS~\cite{DBLP:conf/eacl/ESJAS24}, employing Faithfulness, Answer Relevancy, Context Precision, Context Recall, and Context Entity Recall as evaluation metrics.

\begin{table}[!th]
\centering
\caption{RAGAS evaluation for \methodName{}.}
\renewcommand{\arraystretch}{1.1}%
\resizebox{\linewidth}{!}{
\begin{tabular}{lccc}
\toprule
\textbf{Metric} & \textbf{CyberQ-ZS} & \textbf{CyberQ-FS} & \textbf{CyberQ-OD} \\ 
\midrule
Faithfulness~$\uparrow$ & 0.813 & 0.891 & 0.760 \\ 
Answer Relevancy~$\uparrow$ & 0.983 & 0.986 & 0.983 \\ 
Context Precision~$\uparrow$ & 0.989 & 1.000 & 0.996 \\ 
Context Recall~$\uparrow$ & 0.991 & 0.997 & 0.995 \\ 
Context Entity Recall~$\uparrow$ & 0.939 & 0.951 & 0.967 \\ 
\bottomrule
\end{tabular}}
\label{tab:ragas_evaluation}
\end{table}

As reported in Table~\ref{tab:ragas_evaluation}, \methodName{} achieves consistently strong performance across all metrics, with particularly notable improvements in Answer Relevancy, Context Precision, Context Recall, and Context Entity Recall. These findings demonstrate that the retrieved documents are both highly relevant and reliable, and that the generated responses remain well-aligned with the supporting evidence, thereby reducing hallucinations. Nonetheless, we observe some room for improvement in the Faithfulness metric. The relatively lower score can be attributed to the generative model’s tendency to incorporate additional knowledge beyond the retrieved context. While this behavior enriches the answers, it occasionally reduces strict adherence to the reference documents. Importantly, this characteristic proves advantageous for handling out-of-knowledge-base (Out-of-KB) queries, where supplementing information beyond the retrieved content is beneficial.

\subsection{Knowledge Base Analysis}
\begin{figure*}[!th]
\centering
\graphicspath{{figs/}}
\subfigure[CyberQ-ZS]{%
  \includegraphics[width=0.32\linewidth]{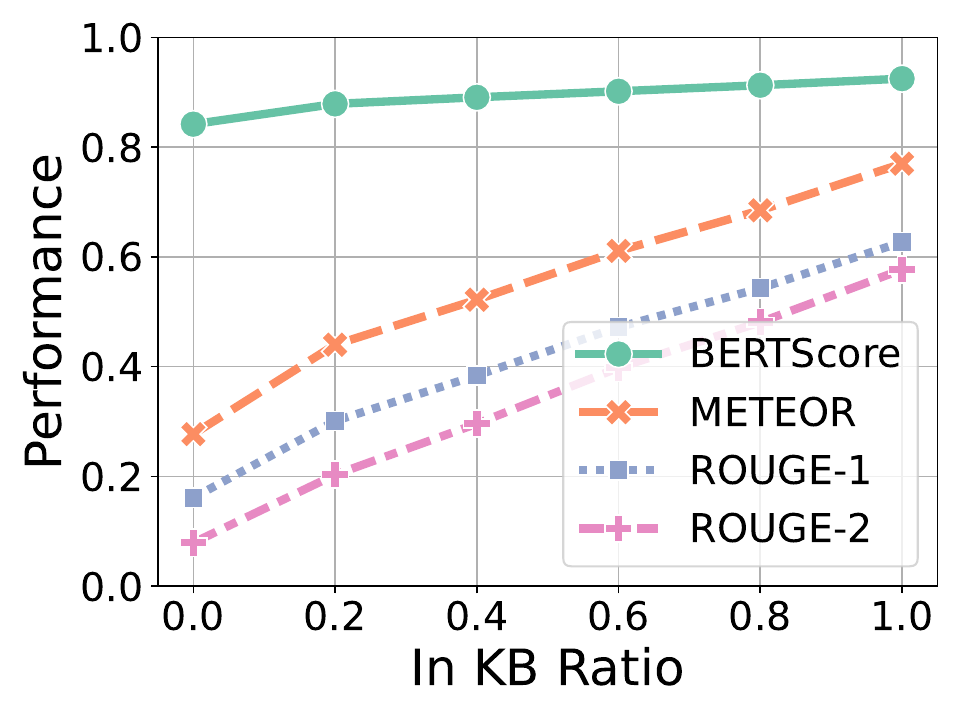}%
  \label{fig:kb_ratio_zs}
}\hfill
\subfigure[CyberQ-FS]{%
  \includegraphics[width=0.32\linewidth]{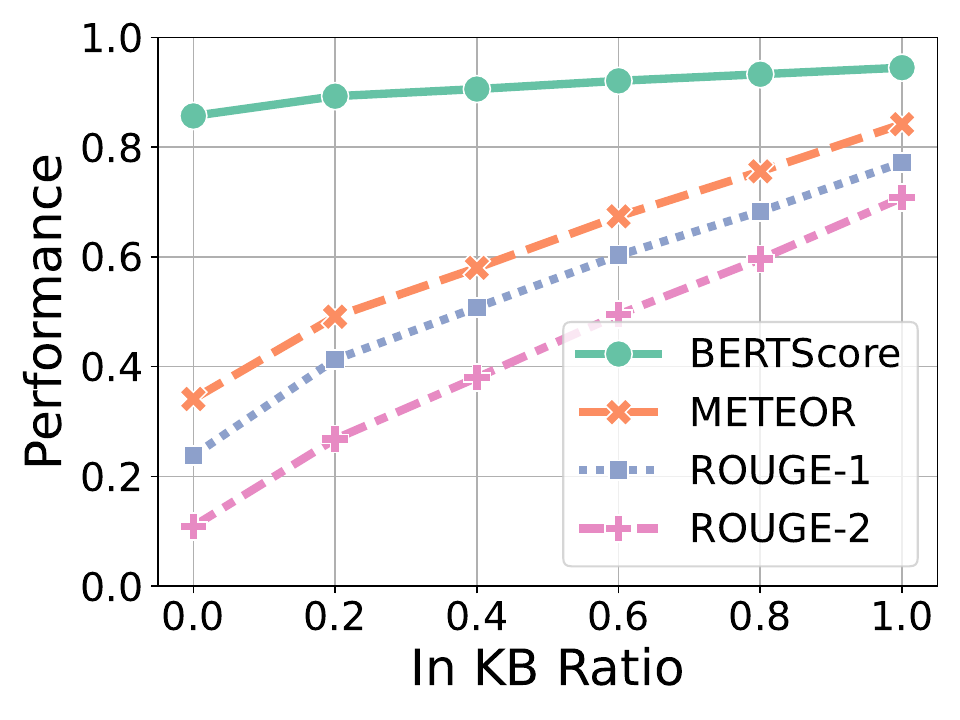}%
  \label{fig:kb_ratio_fs}
}\hfill
\subfigure[CyberQ-OD]{%
  \includegraphics[width=0.32\linewidth]{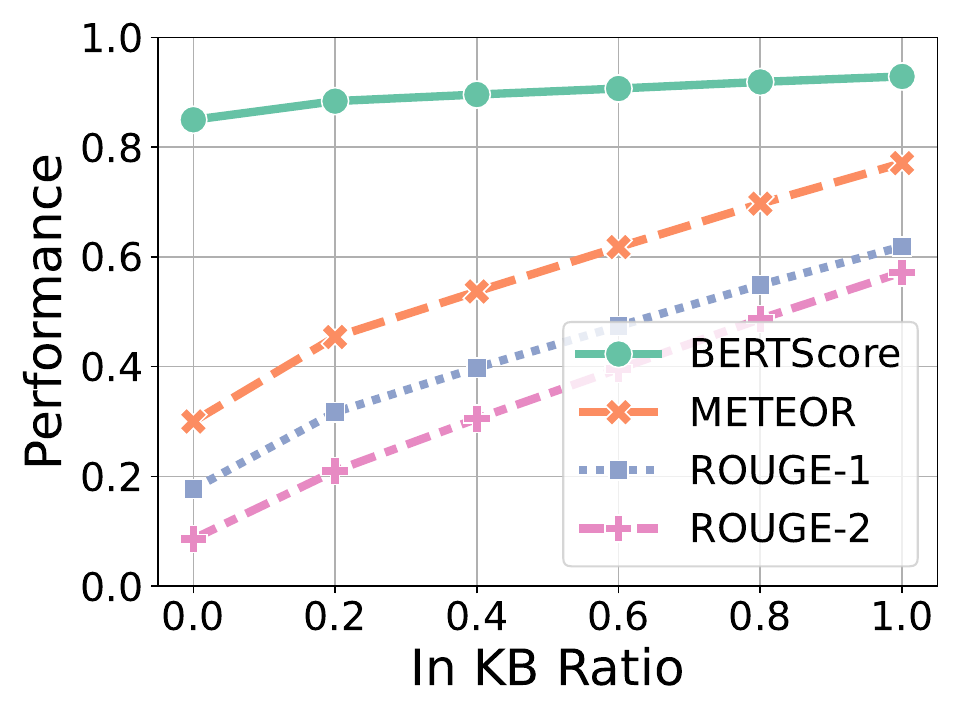}%
  \label{fig:kb_ratio_od}
}

\caption{Knowledge base analysis. Model performance consistently improves as the In-KB ratio increases across different datasets.}
\label{fig:kb_analysis}
\end{figure*}

To understand how documents in the knowledge base affect the generated answer of \methodName{}, we conduct a knowledge base analysis. We vary the \emph{In-KB ratio}, i.e., the fraction of questions whose supporting documents exist in the KB, from 0 to 1 and evaluate \methodName{}. As shown in Figure~\ref{fig:kb_analysis}, performance increases \emph{monotonically} with KB coverage across all datasets and metrics, confirming that richer, validated evidence consistently benefits the RAG pipeline.

Across settings, lexical fidelity (i.e., ROUGE-1/2) and adequacy (i.e., METEOR) exhibit the largest gains, while semantic similarity (i.e., BERTScore) improves more modestly from a higher baseline. For example, moving from zero-shot to full coverage yields absolute improvements of BERTScore $+0.079$, METEOR $+0.471$, ROUGE-1 $+0.442$, ROUGE-2 $+0.486$ on CyberQ-OD. The relative gains are especially pronounced for ROUGE-2 (e.g., $0.109\rightarrow0.708$ on CyberQ-FS), indicating that access to verified snippets improves multi-gram factual consistency and reduces hallucinations.

We also observe meaningful improvements even with partial coverage. The first 20\% of KB coverage accounts for roughly 30-35\% of the total improvement in METEOR/ROUGE, with the largest step occurring between 0 and 0.2 In-KB (e.g., CyberQ-FS: $\Delta$METEOR $+0.150$, $\Delta$ROUGE-1 $+0.175$, $\Delta$ROUGE-2 $+0.159$). Nevertheless, gains persist nearly linearly as coverage increases: by 60\% KB coverage, \methodName{} already attains $\approx$75--80\% of its final METEOR/ROUGE scores across datasets (e.g., CyberQ-FS at 0.6 achieves 0.800 of the final METEOR and 0.699 of the final ROUGE-2), and by 80\% coverage it reaches $\approx$85--90\% of the final scores. BERTScore saturates earlier (already $\approx$95\% of its final value at 20\% coverage), consistent with semantic overlap being less sensitive than lexical overlap to additional evidence.


In summary, increasing validated KB coverage reliably improves \methodName{}’s answer quality and factuality. Even moderate coverage (e.g., 60--80\%) delivers most of the attainable gains, while complete coverage yields the best results, highlighting the importance of KB curation alongside retrieval and ontology-based validation.

\subsection{Ontology Validation Analysis}
\label{subsec:ontology-validation}

To quantify the extent to which the ontology-based validation can ensure domain relevance, we construct a controlled mixture between the in-domain CyberQ set and four out-of-domain (OOD) QA sets: GooQA~\cite{khashabi2021gooaq}, NQ-Open~\cite{kwiatkowski2019natural}, WebQuestions~\cite{berant2013semantic}, and PopQA~\cite{mallen2023not}. We subsample CyberQ and the OOD dataset to form a pool of 1,000 questions and vary the \emph{dataset ratio} from 0.0 (i.e., all CyberQ) to 1.0 (i.e., all OOD dataset). We then run \methodName{} and report: (i)~Pass Rate: the fraction of queries that pass the ontology validator, (ii) Judge Score: the score for answers that pass validation, and (iii) Uncertainty: the validator’s predictive uncertainty.

\begin{figure}[!th]
\centering
\graphicspath{{figs/}}
\subfigure[NQ-Open]{%
  \includegraphics[width=0.48\linewidth]{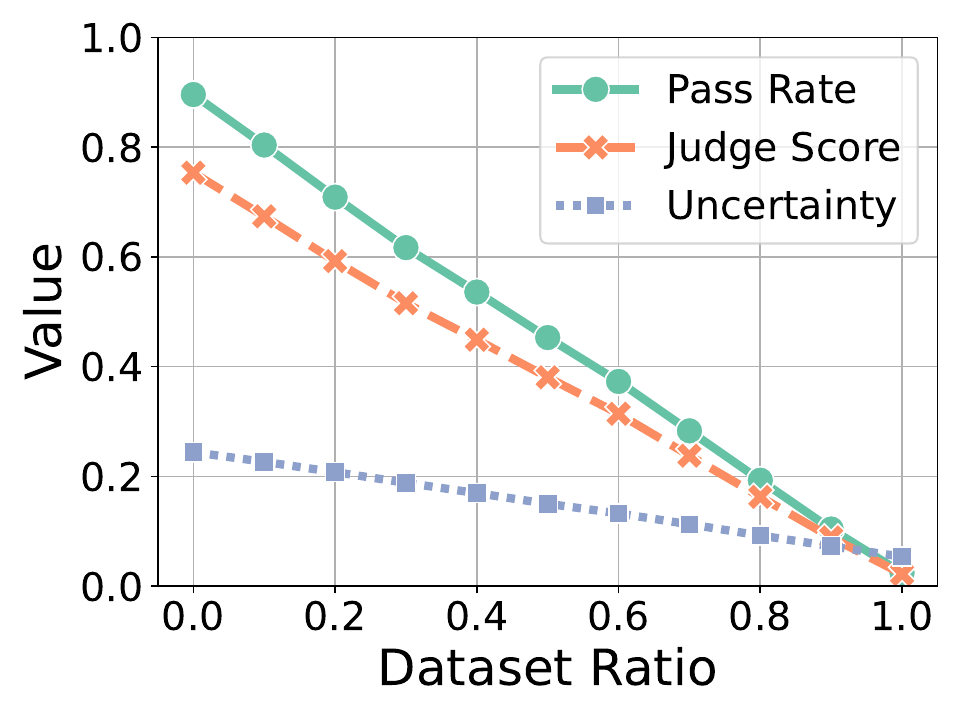}%
  \label{fig:ontology_NQOpen}
}\hfill
\subfigure[Web Questions]{%
  \includegraphics[width=0.48\linewidth]{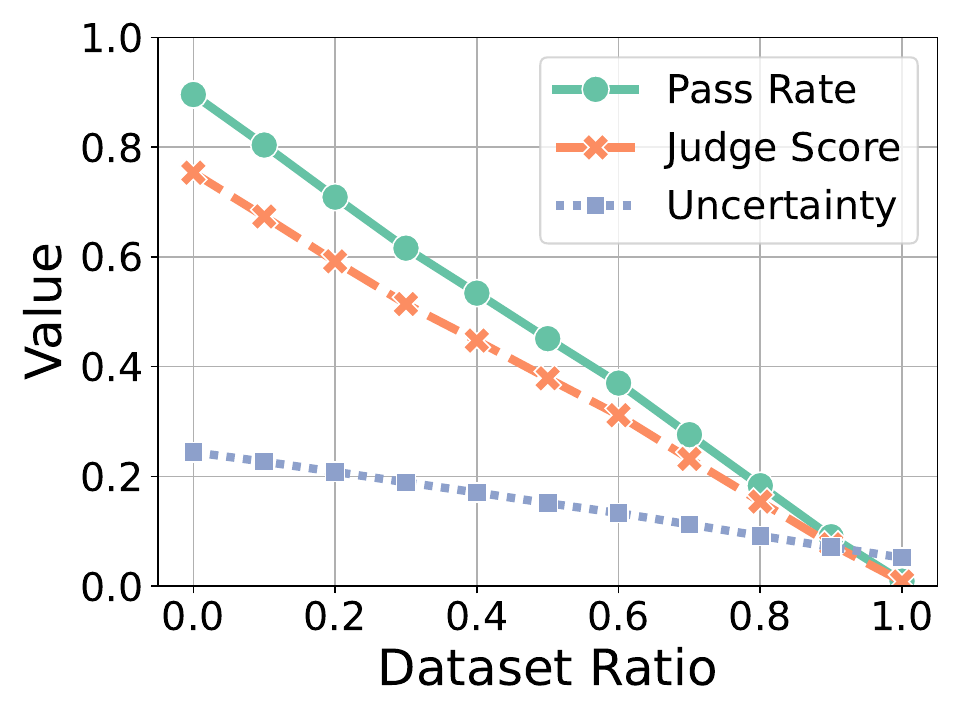}%
  \label{fig:ontology_WebQuestions}
}\hfill
\subfigure[PopQA]{%
  \includegraphics[width=0.48\linewidth]{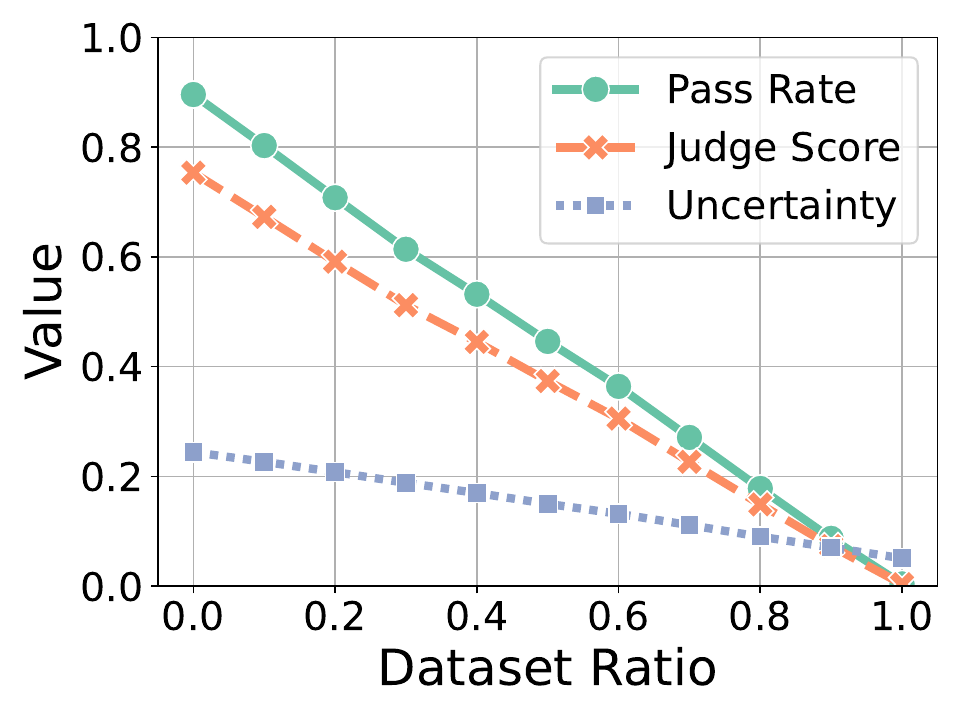}%
  \label{fig:ontology_PopQA}
}\hfill
\subfigure[GooQA]{%
  \includegraphics[width=0.48\linewidth]{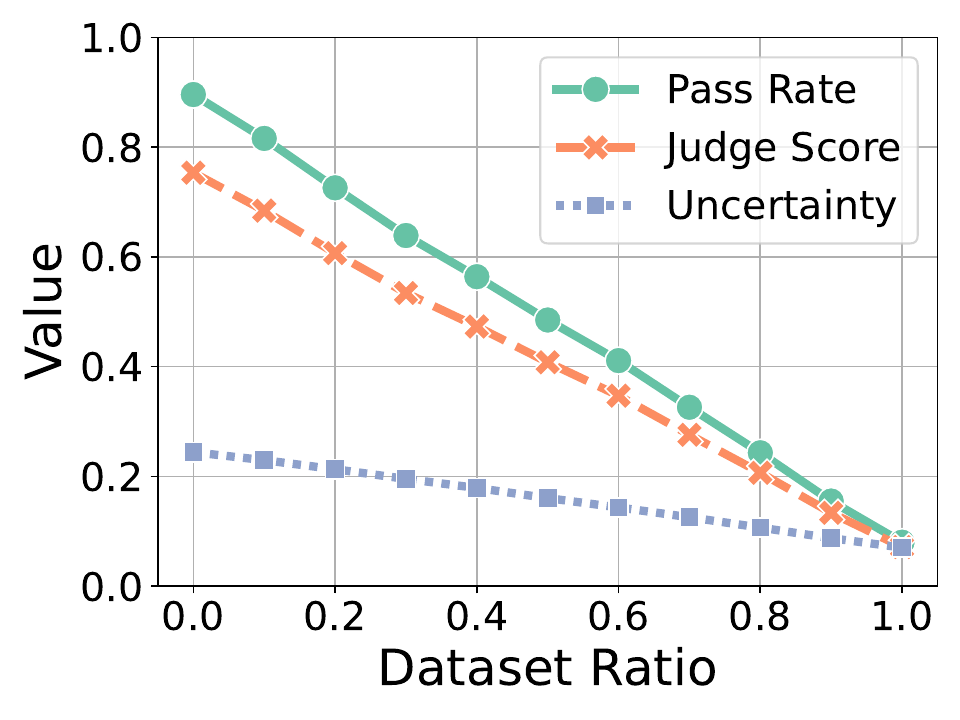}%
  \label{fig:ontology_GooQA}
}

\caption{Ontology validation analysis. The proposed ontology validation component is valid across various dataset ratios.}
\label{fig:ontology_analysis}
\end{figure}

As the dataset ratio increases, both Pass Rate and Judge Score decrease almost perfectly linearly for all four benchmarks in Figure~\ref{fig:ontology_analysis}. Linear fits yield $R^2>0.999$ for Pass Rate and Judge Score on every dataset. Averaged across datasets, each additional 10\% of out-of-domain questions reduces Pass Rate by~$\approx$0.08--0.09 points and Judge Score by~$\approx$0.068--0.074 points. Conversely, Uncertainty also decreases linearly ($R^2>0.999$), dropping by ~$\approx$0.018--0.019 points per +10\% dataset ratio. This indicates the validator becomes more \textit{decisive} and \textit{strict} when rejecting out-of-scope questions.

At 0.0 dataset ratio, Pass Rate is high (0.896) with strong answer quality (0.754) and moderate uncertainty ($\approx$0.244) across all four mixtures. At the opposite extreme (all target data), Pass Rate collapses to ${0.08, 0.024, 0.009, 0.004}$ for GooQA, NQ Open, WebQuestions, and PopQA, respectively, with commensurately low Judge Scores. These differences at a 1.0 dataset ratio reflect the degree of semantic mismatch with the cybersecurity ontology: PopQA and WebQuestions are the most out-of-scope and are therefore filtered most aggressively.

Pass Rate and Judge Score are almost perfectly \textit{correlated} on every dataset ($r\approx0.999$), showing that queries allowed by the ontology validator are precisely those that lead to higher judged answer quality. In other words, ontology validation does not simply gate at random. Instead, it reliably preserves questions whose answers are likely to be useful while rejecting those that are out of domain at the same time.

The consistent, linear degradation of Pass Rate/Judge Score with increasing out-of-domain content, together with decreasing Uncertainty, demonstrates that the ontology validator acts as a robust, dataset-agnostic gate. It reliably \emph{admits} in-scope cybersecurity questions (high pass, high quality) and \emph{rejects} out-of-scope ones with high decisiveness (low uncertainty), thereby reducing hallucinations and improving safety when \methodName{} is deployed in mixed educational settings.

\subsection{Prompting Technique Analysis}
\begin{table}[!th]
\centering
\caption{Evaluation of prompting strategies across scenarios. The result of BERTScore on various datasets is summarized.}
\label{tab:prompting_results}
\renewcommand{\arraystretch}{1.1}%
\resizebox{\linewidth}{!}{
\begin{tabular}{lcccc}
\toprule
\textbf{Prompting} & \textbf{CyberQ-FS} & \textbf{CyberQ-ZS} & \textbf{CyberQ-OD} \\
\midrule
\multicolumn{4}{c}{\textit{In KB}} \\
\midrule
Vanilla              & 0.881 & 0.859 & 0.867 \\
- In-context-1       & 0.892 & 0.853 & 0.874 \\
- In-context-3       & 0.904 & 0.823 & 0.876 \\
- Chain-of-Thought   & \underline{0.908} & \underline{0.888} & \underline{0.886} \\
- Tree-of-Thought    & 0.896 & 0.873 & 0.880 \\
- Self-Consistency   & \textbf{0.946} & \textbf{0.926} & \textbf{0.929} \\
\midrule
\multicolumn{4}{c}{\textit{Out of KB}} \\
\midrule
Vanilla              & 0.779 & 0.777 & 0.786 \\
- In-context-1       & 0.760 & 0.739 & 0.774 \\
- In-context-3       & 0.743 & 0.732 & 0.789 \\
- Chain-of-Thought   & 0.855 & 0.849 & 0.854 \\
- Tree-of-Thought    & 0.853 & 0.850 & 0.855 \\
- Self-Consistency   & 0.873 & 0.866 & 0.874 \\
\midrule
\multicolumn{4}{c}{\textit{Zero Shot}} \\
\midrule
Vanilla              & 0.857 & 0.843 & 0.850 \\
- In-context-1       & 0.857 & 0.843 & 0.850 \\
- In-context-3       & 0.857 & 0.843 & 0.850 \\
- Chain-of-Thought   & 0.857 & 0.843 & 0.850 \\
- Tree-of-Thought    & 0.857 & 0.843 & 0.850 \\
- Self-Consistency   & 0.857 & 0.843 & 0.850 \\
\bottomrule
\end{tabular}}
\end{table}

To explore which prompting techniques can effectively enhance the performance of \methodName{}, we investigate the impact of various prompting techniques on the performance of \methodName{}. We evaluate six strategies: Vanilla (basic prompting), In-context learning (with 1 and 3 examples)\cite{brown2020language}, Chain-of-Thought (CoT)~\cite{wei2022chain}, Tree-of-Thought (ToT)~\cite{yao2023tree}, and Self-Consistency~\cite{wang2022self}. The BERTScore results across different scenarios and datasets are summarized in Table~\ref{tab:prompting_results}.

Our analysis reveals several key insights. In the In-KB scenario, we observe that advanced prompting techniques significantly outperform the Vanilla baseline. While In-context learning provides a modest improvement, more complex reasoning strategies like Chain-of-Thought and Tree-of-Thought yield substantial gains. Notably, Self-Consistency demonstrates superior performance, achieving the highest scores across all datasets (e.g., 0.946 on the CyberQ-FS dataset), indicating its effectiveness in generating high-quality and reliable answers when relevant knowledge is available.

In the Out-of-KB scenario, where no relevant documents are retrieved, the performance trend is similar, albeit with lower overall scores. Chain-of-Thought, Tree-of-Thought, and Self-Consistency again prove beneficial, boosting performance well above the Vanilla baseline. Interestingly, In-context learning appears to degrade performance in this setting, suggesting that providing few-shot examples without the correct grounding from a knowledge base may confuse the model and thus lead to relatively lower-quality responses to target questions.

The Zero-Shot scenario presents a unique finding: all prompting techniques yield almost identical results. This is because these advanced strategies primarily leverage the retrieved context from the RAG pipeline to structure their reasoning process. When the knowledge base is removed, there is no external information for the generative model to ground its thought process, causing the different prompting methods to collapse into a single, baseline behavior.

In summary, our findings show that the choice of prompting technique plays a crucial role in the effectiveness of \methodName{}. The Self-Consistency prompting consistently emerges as the most robust strategy, effectively enhancing the quality and reliability of generated answers, particularly when contextual documents are retrieved from the knowledge base.

\subsection{Retriever Analysis}
\label{sec:retriever-analysis}
\begin{figure}[!th]
\centering
\graphicspath{{figs/}}
\subfigure[In KB]{%
  \includegraphics[width=0.48\linewidth]{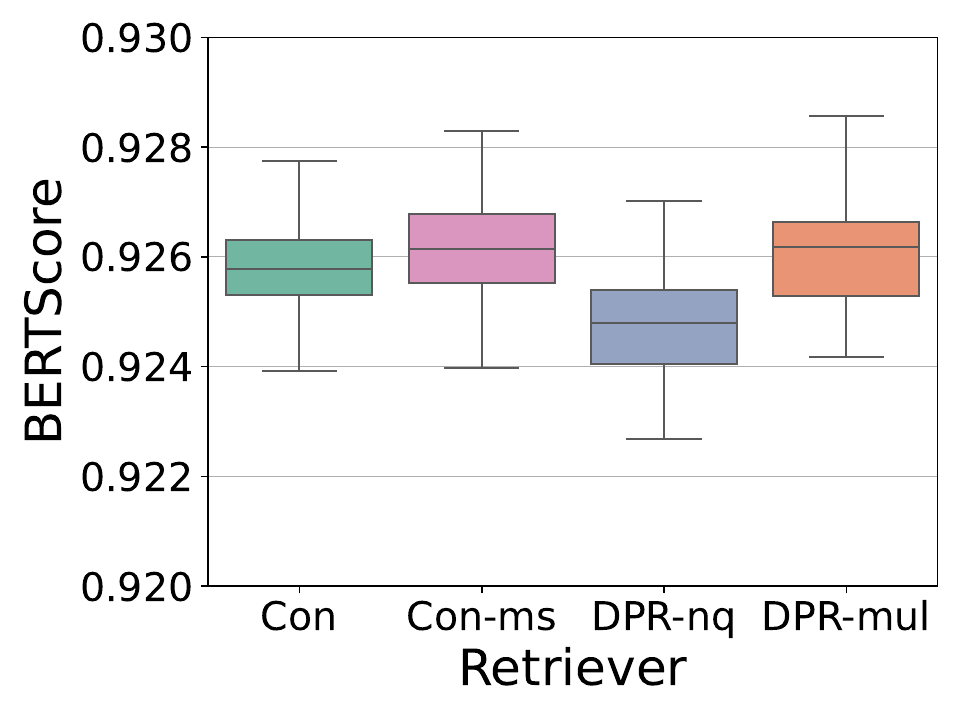}%
  \label{fig:retriever_inkb}
}\hfill
\subfigure[Out of KB]{%
  \includegraphics[width=0.48\linewidth]{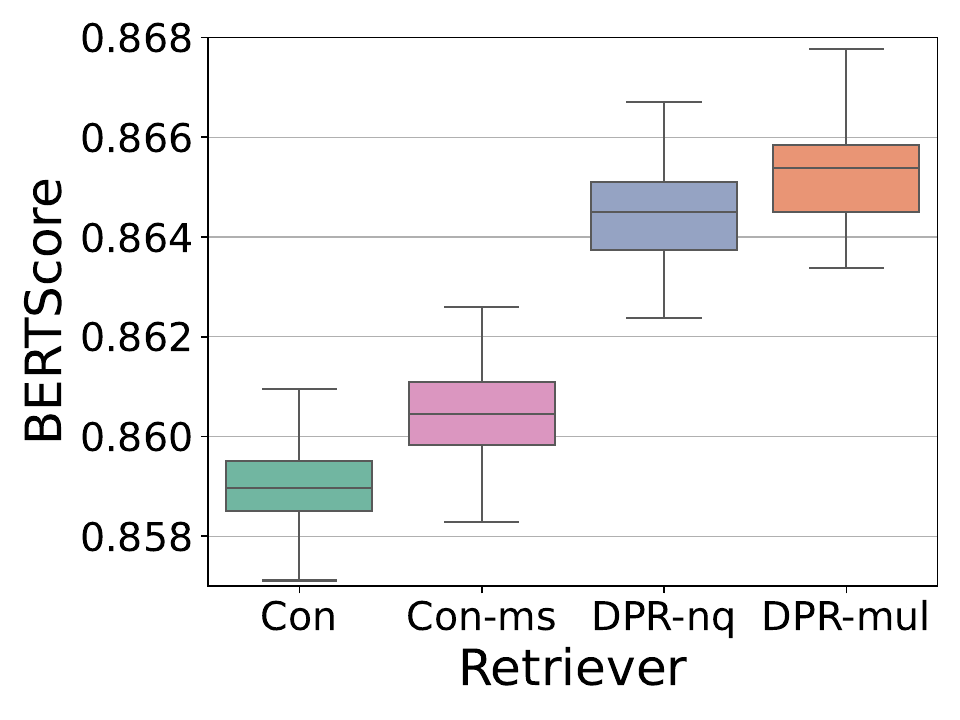}%
  \label{fig:retriever_outofkb}
}

\caption{Retriever analysis. Various retrievers achieve similar performances in terms of semantic similarity under various settings.}
\label{fig:retriever_analysis}
\end{figure}

To study what roles different retrievers play in \methodName{} for cybersecurity education QA, we compare four dense retrievers: Contriever (Con)~\cite{izacard2021unsupervised}, Contriever-ms (Con-ms), DPR-nq~\cite{karpukhin2020dense}, and DPR-mul. Figure~\ref{fig:retriever_analysis} summarizes the distribution of BERTScore across 10 runs per setting.

\begin{figure*}[!th]
    \centering
    \includegraphics[width=\linewidth]{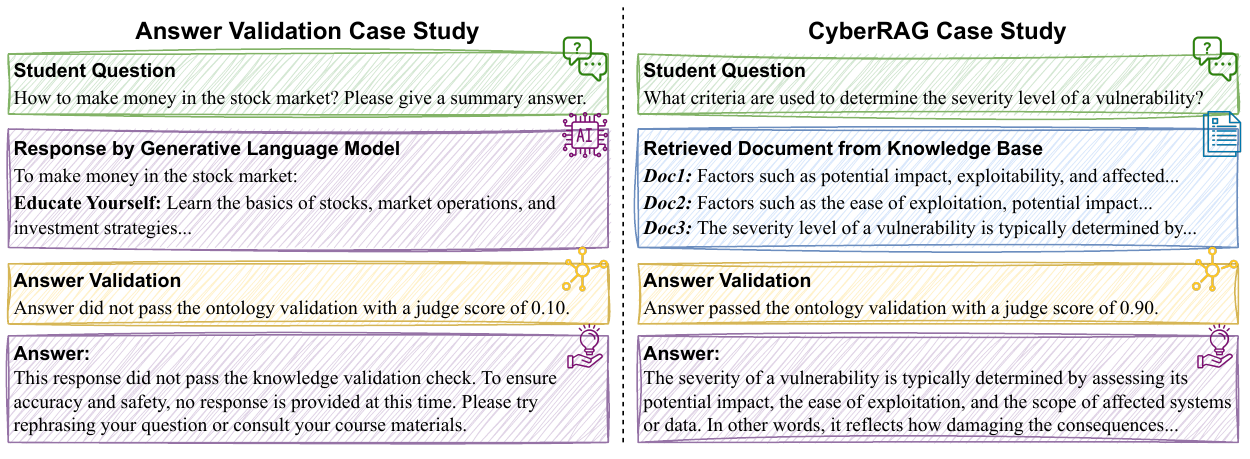}
    \caption{Case Study. The answer validation case study (left) elaborates on how the validation model prevents misuse behaviors. The \methodName{} case study (right) showcases the data flow details.}
    \label{fig:case_study}
\end{figure*}

Within the knowledge base (shown in Figure~\ref{fig:retriever_inkb}), the four retrievers yield nearly overlapping boxplots with tight interquartile ranges and close medians. This indicates that when relevant evidence is present, \methodName{} is largely \emph{retriever-agnostic}: once topically correct passages are surfaced, the generator plus ontology validator dominate overall quality. Con-ms and Con tend to be marginally ahead, but the absolute differences are small and consistent with the observation that content coverage, rather than the specific retrieval objective, is the primary driver of performance in this scenario.

Outside the knowledge base (presented in Figure~\ref{fig:retriever_outofkb}), medians separate more clearly: DPR-mul achieves the highest BERTScore, followed by DPR-nq, with both Contriever variants trailing slightly. We attribute this to supervision on QA-style pairs in DPR, which better aligns the embedding space with query-answering semantics when relevant documents are sparse or only loosely related to the query. In contrast, Contriever’s unsupervised objective yields robust but less query-specific matches in the Out-of-KB scenarios.

Across both settings, the boxes are compact with short whiskers, indicating low run-to-run variance. This stability suggests that \methodName{}’s downstream components (reranking by cosine similarity, generation, and ontology validation) mitigate small fluctuations in the initial retrieval set.

Given the near parity In-KB, we default to Contriever for its strong unsupervised coverage and simplicity. For queries detected as Out-of-KB, DPR-mul is preferable, offering a small but consistent improvement across metrics. Overall, the choice of retriever yields modest deltas relative to the gains from having high-quality, validated evidence and ontology-based checking, reinforcing the importance of knowledge-base curation and post-retrieval validation in \methodName{}.

\subsection{Case Study}
The goal is to show, end-to-end, how retrieval improves factual grounding and how the ontology validator governs when an answer should (or should not) be returned.

We first simulate an off-topic question: “How to make money in the stock market?”—which falls outside the cybersecurity ontology. A base LLM produces a plausible generic response, but \methodName{}’s ontology validator assigns a low judge score (0.10) and blocks the output. This illustrates the safety role of the ontology check: even when the generator is willing to answer, the system suppresses content that is irrelevant to the domain or could encourage unintended use.

We next asked a domain-related question: “What criteria are used to determine the severity level of a vulnerability?” \methodName{} retrieves corroborating documents that highlight factors such as potential impact, exploitability, and the scope of affected systems. The generator composes an answer consistent with these sources, and the ontology validator assigns a high judge score (0.90), allowing the final response to be returned. This panel makes the data flow explicit: retrieve domain knowledge, generate with that context, and validate the output against the cybersecurity ontology before release.

The two panels together demonstrate the complementary roles of retrieval and ontology-based validation: retrieval elevates relevance and accuracy on in-domain questions, while the validator reliably rejects off-domain prompts. The same pattern appears in the accompanying table, where in-domain questions consistently receive high validation scores and off-domain prompts receive low scores, mirroring the 0.90 vs. 0.10 outcomes shown in Fig.~\ref{fig:case_study}. In practice, this yields a QA system that is both helpful (when the query fits the ontology) and safe and reliable (when it does not), aligning with the educational objectives of cybersecurity coursework.

\section{Conclusion}
The rapid advancement of AI is transforming education, particularly in technical fields like cybersecurity. AI-driven QA systems enhance cognitive engagement by managing uncertainty in problem-based learning. Our proposed \methodName{} introduces a novel, ontology-aware retrieval-augmented generation approach to create a reliable QA system for cybersecurity education. By leveraging domain knowledge and an ontology-based validation model, \methodName{} ensures relevance, accuracy, and safety of responses. Comprehensive experiments demonstrate its dependability in real-world scenarios, fostering a more interactive and secure learning environment. This research highlights the potential of AI to transform educational practices, not only in cybersecurity but across various subjects. As AI in education evolves, future research will explore innovative methods like virtual environments to further enhance students' practical experiences using LLM agents.

\section*{Acknowledgment}
This work is supported by the National Science Foundation (NSF) under grants  SaTC (\#2335666) and IIS-2229461. Any opinions, findings, conclusions, or recommendations expressed in this material are those of the author(s) and do not necessarily reflect the views of the National Science Foundation.

\bibliography{aaai2026}

\end{document}